\documentclass[preprint,amsmath,amssymb,pra,groupedaddress]{revtex4-1}

\usepackage{graphicx}
\usepackage{dcolumn}
\usepackage{bm}
\usepackage{multirow}

\begin{document}

\preprint{APS/123-QED}

\title{Spin Damping\\in an RF Atomic Magnetometer}

\author{Orang Alem}
 \email{oranga@gmail.com}
 \affiliation{Department of Physics and Astronomy, George Mason University, Fairfax, Virginia, 22030, USA.}

\author{Karen L. Sauer}
 \affiliation{Department of Physics and Astronomy, George Mason University, Fairfax, Virginia, 22030, USA.}

\author{Mike V. Romalis}
 \affiliation{Department of Physics, Princeton University, Princeton, New Jersey, 08544, USA.}

\date{\today}

\begin{abstract}

Under negative feedback, the quality factor $Q$ of a radio-frequency magnetometer can be decreased by more than two orders of magnitude, so that any initial perturbation of the polarized spin system can be rapidly damped, preparing the magnetometer for detection of the desired signal. We find that noise is also suppressed under such spin-damping, with a characteristic spectral response corresponding to the type of noise; therefore magnetic, photon-shot, and spin-projection noise can be measured distinctly.  While the suppression of resonant photon-shot noise implies the closed-loop production of polarization-squeezed light, the suppression of resonant spin-projection noise does not imply spin-squeezing, rather simply the broadening of the noise spectrum with $Q$. Furthermore, the application of spin-damping during phase-sensitive detection suppresses both signal and noise in such a way as to increase the sensitivity bandwidth.  We demonstrate a three-fold increase in the magnetometer's bandwidth while maintaining 0.3~fT/$\sqrt{\mathrm{Hz}}$ sensitivity.

\end{abstract}


\maketitle

\section{\label{Introduction}Introduction}

Ideally, a sensor of radio-frequency magnetic fields is sensitive over a broad bandwidth and has a fast recovery time.  The last requirement is particularly important when pulsed excitation is used to create the detected signal, as in detection by nuclear magnetic or nuclear quadrupole resonance.  In conventional magnetic resonance detection a coil of wire is used both to excite the sample and detect the resulting signal.  Hoult in 1979 successfully applied negative feedback to damp such a probe so that the recovery time was reduced but the signal-to-noise ratio remained the same during data acquisition \cite{Hoult-1979}.  While Hoult used negative feedback to change the impedance of the detection circuit and thus the quality factor $Q$ of the probe, more recent work has focused on using negative feedback to generate emf that directly opposes the emf already in the coil \cite{Nacher-2011}; both a decrease in recovery time and an increase in signal bandwidth without the loss of signal-to-noise ratio was observed.  The latter usage of negative feedback is close in principle to the damping described here for an atomic system.

Recently, atomic magnetometers using optically-pumped alkali atoms have been shown to be more sensitive to radio-frequency magnetic fields than standard coil detection \cite{Savukov-Seltzer-Romalis-2007}, particularly at low frequencies as is needed for low-field magnetic resonance \cite{Savukov-Romalis-2005,Ledbetter-Pines-etal-2008,Xu-Pines-2008,Savukov-Espy-etal-2011,Oida-etal-2012} or nuclear quadrupole resonance \cite{Lee-Alem-etal-2006}.  Sensitivities as low as 0.2~fT/$\sqrt{\mathrm{Hz}}$ are gained at the expense of operating with a relatively narrow signal bandwidth, on the order of a half a kHz, or a correspondingly long alkali spin-spin relaxation time $T_2$ of about a millisecond \cite{Lee-Alem-etal-2006}.  A long $T_2$ also contributes towards long recovery times.  Therefore for short-lived signals, or those applications which require good sensitivity over a large bandwidth, this time constant can be prohibitively long.

One way to broaden the magnetometer's sensitivity is to use continuous quantum non-demolition measurements on a magnetometer limited predominantly by spin projection noise, as was demonstrated by Ref.~\cite{Shah-Vasilakis-Romalis-2010} for a scalar magnetometer.  With a spin-polarization of 1~$\%$, Ref.~\cite{Shah-Vasilakis-Romalis-2010} achieved a four fold increase in sensitivity bandwidth while maintaining a sensitivity of 22~fT/$\sqrt{\mathrm{Hz}}$; with higher polarization, they estimate that a sensitivity $\sim 0.6$ fT/$\sqrt{\mathrm{Hz}}$ and a two fold increase in bandwidth can be realized.  The magnetometer presented in this paper has a sensitivity of $\sim$0.3~fT/$\sqrt{\mathrm{Hz}}$ and is dominated by environmental magnetic and photon shot noise.  With these dominate noise sources, we find another way that sensitivity bandwidth can be broadened without significant loss of sensitivity - through negative feedback.

To implement negative feedback, the AC signal from the magnetometer is converted to a magnetic field and applied back to the magnetometer so as to damp out the transverse atomic polarization responsible for the signal; the basic schematic is shown in Fig.~\ref{Pp-MagSDSchematic}(b).  In analogy to $Q$-damping, we term this spin-damping. We will show that spin-damping lowers the effective $T_2$ of the spin system.  The strength of the damping can be easily controlled by the gain/attenuation of the signal which is fed back, permitting rapid changes in the effective $T_2$.  We demonstrate that spin-damping can be used to gain a fast recovery time for the magnetometer and, for phase-sensitive detection, can be used to increase the detector bandwidth with negligible loss of detector sensitivity.

The idea of using negative feedback to push atomic spin system back into alignment was originally proposed by researchers at Caltech \cite{Stockton-Mabuchi-etal-2004}. As in our system, Faraday rotation and a balanced polarimeter is used as a measure of the spin polarization along the probe laser beam direction, but unlike in our system, the signal and therefore the feedback field is inherently DC.  Although they were unable to demonstrate their initial goal of suppressing spin-projection noise below the standard quantum limit \cite{Geremia-etal-2004,Geremia-retraction-2008}, ie. spin-squeezing \cite{Wasilewski-Polzik-etal-2010}, they did show that negative feedback impacted the measured noise of the system.  We will demonstrate that for our system under damping, while the total integrated noise power is reduced for magnetic and photon shot noise, it remains the same for spin-projection noise and is therefore not an example of spin-squeezing.  Rather the spectrum of spin-projection noise is broadened according to the effective $T_2$ under damping.  Because the different sources of noise behave distinctly from one another under damping, spin-damping permits a way to measure the spin-projection noise in a spin system, even if it is much smaller than the other sources of noise.

\section{\label{Experimental Setup}Experimental Setup}

\begin{figure}[h!]
\centerline{\includegraphics[width=6in]{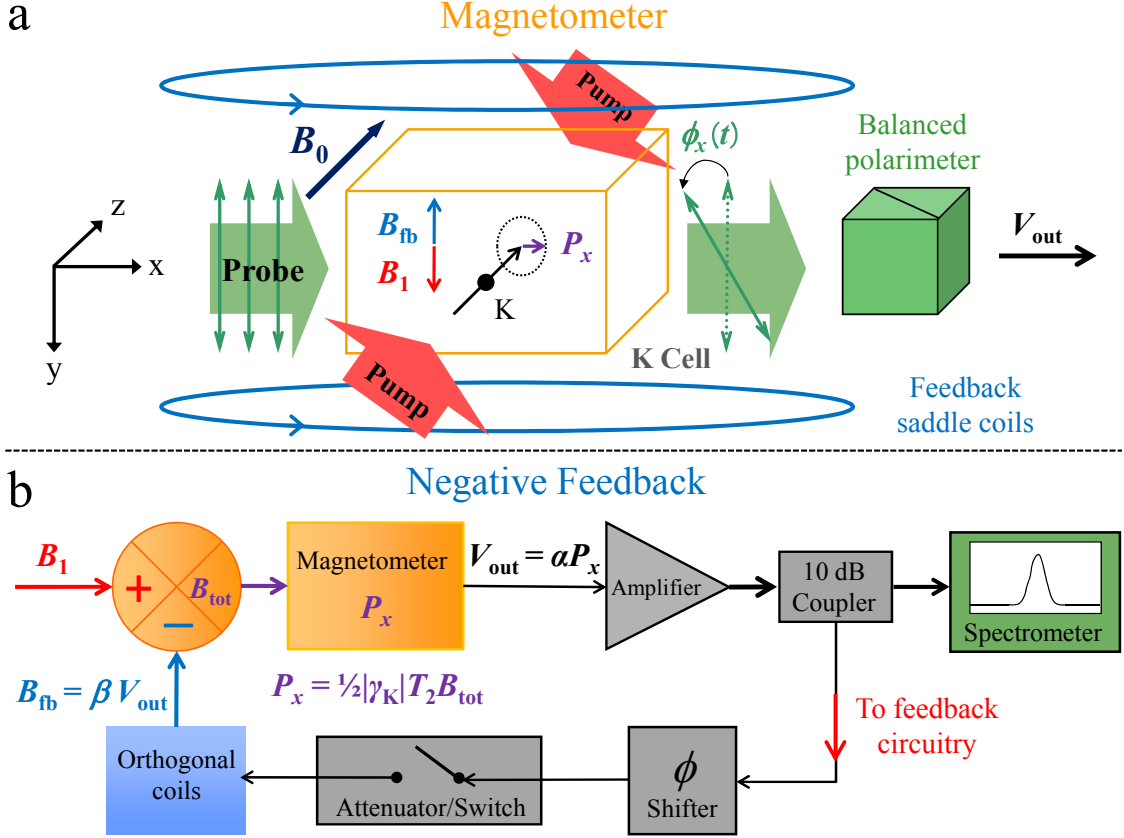}}
\caption[Spin-damping schematic]{Experimental set-up (a) and schematic (b) of the spin-damping mechanism. A perturbing magnetic field $B_{\rm 1}$ sets the optically pumped K atoms precessing around $B_0$. The resulting transverse atomic polarization $P_x$ rotates the probe beam's polarization, which is detected and converted to an electrical signal $V_{\rm out}$ by a balanced polarimeter. Part of this electric signal is phase corrected and fed back through electromagnetic saddle coils to produce a damping field $B_{\rm fb}$ that is anti-parallel to $B_{\rm 1}$, resulting in an active damping of the K transverse polarization. The strength of the damping is characterized by the open loop gain, $DF= \alpha \beta P_z \gamma_K/2$.}
\label{Pp-MagSDSchematic}
\end{figure}

Our basic experimental set-up, as shown in Fig.~\ref{Pp-MagSDSchematic}(a), is similar to the scheme used in Ref.~\cite{Lee-Alem-etal-2006} to detect NQR signals of ammonium nitrate at its characteristic frequency of 423~kHz. A set of Barker coils \cite{Robinson-1983}, inside a triple set of mu metal magnetic shields and an aluminum RF shield, is used to apply a small static magnetic field of $B_0 = 60~\rm \mu T$$\hat{z}$ to tune the resonance of the magnetometer to 423~kHz. In addition, a set of three coils serve to correct for first-order stray gradients of the static field in $\hat{z}$ \cite{Golay-1958,Suits-Wilken-1989} and a pair of saddle coils \cite{Hoult-1976} produces fields in $\hat{x}$ and $\hat{y}$, orthogonal to the static field $B_0$. The saddle coils not only serve to produce static and RF fields of known strength for calibrating the magnetometer, but also serve as an integral part of the feedback mechanism used for spin-damping. As shown in Fig.~\ref{Pp-MagSDSchematic}, part of the RF signal from the magnetometer is applied to one of the saddle coils. The phase and amplitude of this signal is adjusted to produce negative feedback of a known strength, to produce a damping field $B_{\rm fb}$.

At the center of these electromagnetic coils sits a $4 \times 4 \times 6 ~\rm cm$ K vapor cell. A non-magnetic hot air oven, with four optical windows, directly surrounds the cell and keeps it at $180^\circ\pm1^\circ \rm C$. In principle, the temperature of the cell sets the K number density in the vapor \cite{Alcock-1984}. However, due to the interaction of the alkali metal with the pyrex cell walls \cite{Doremus-1977,Laux-Schiltz-1980,Zhao-Wu-Lai-2001}, the number density of the cell is reduced and varies from cell to cell \cite{Jau-Kuzma-Happer-2002}. For the data presented in this paper the K vapor density of $4\times10^{13}~\rm cm^{-3}$ is measured using the resonant linewidth at 100~kHz in the limit of low pump and probe power where the broadening is dominated by K spin-exchange collisions \cite{Happer-Tam-1977,Savukov-Sauer-etal-2005,Vasilakis-Shah-Romalis-2011}. In addition to the K droplets, the cell contains $\sim 650~\rm torr$ of He, to slow the diffusion to the wall, and $\sim 60~\rm torr$ of $\rm N_2$, to serve as a quenching gas.

The cell is illuminated with two tunable single-mode continuous-wave diode lasers with a narrow linewidth of $<300~\rm kHz$ at 770~nm \footnote{LION External Cavity continuous-wave tunable diode Lasers in Littman/Metcalf configuration, Sacher Lasertechnik Group, Marburg, Germany, www.sacher-laser.com.} and which provide up to 1~W of pump and probe light. The K vapor is optically pumped by two counter-propagating circularly-polarized pump beams at the K D$_1$ line. This configuration, shown in Fig.~\ref{Pp-MagSDSchematic}(a), provides a relatively uniform K polarization along $\hat{z}$, which can be determined from the response of the magnetometer to a small static magnetic field, applied along the probe direction, and the measured atomic density \cite{Vasilakis-Shah-Romalis-2011}. Typical K polarizations of at least $75\%$ are readily achieved, with higher polarization hindered primarily by K film build-up on the cell walls. A far off-resonance linearly polarized beam passes through the vapor cell and probes the net transverse magnetization in $\hat{x}$. The resulting Faraday rotation is measured with a balanced polarimeter. Typically, the probe power incident on the cell is 30~mW with a wavelength of 769.72~nm.  In practice, the power of the pump beam is then chosen to optimize the signal, so that we operate close to the maximum $T_2$ of the magnetometer.  Under these conditions, the output from the balanced polarimeter gives a magnetometer responsivity of $(V_{\rm out}/B_{\rm RF}) = 0.55~\rm \mu V/fT$ and we observe a resonant linewidth of about 400~Hz, corresponding to a magnetometer $Q$ of approximately 1000.

The magnetometer output $V_{\rm out}$ is amplified by a factor of 10 or more before it is recorded by a Tecmag spectrometer using quadrature detection \footnote{Apollo DSP console, Tecmag Inc., Houston, Texas, www.tecmag.com.}. The sensitivity of the magnetometer in this configuration is fundamentally limited by photon shot noise at $0.1~\rm fT/\sqrt{Hz}$. However, the presence of environmental noise, most probably due to magnetic field noise from the excess K metal within the cell and the wires adjacent to the cell \cite{Alem-2011,Lee-Romalis-2008}, limits the measured sensitivity. Optimal sensitivity of $0.22\pm0.02~\rm fT/\sqrt{Hz}$ was achieved using a cell with a small amount of K which was eventually completely absorbed by the glass \cite{Doremus-1977,Laux-Schiltz-1980}. Before absorption and loss of signal, the number density, interestingly enough, was approximately half that of cells with more K. For the data presented later in this article, we used two cells with noticeably more K, and we refer to them as cell 1 and cell 2. Due to variations in oven assembly, cell 2 developed considerably more film on the optical surfaces, which we believed resulted in a worse sensitivity, $0.37\pm0.03~\rm fT/\sqrt{Hz}$,  and lower polarization, $78\%$, than cell 1, with sensitivity of $0.26\pm0.02~\rm fT/\sqrt{Hz}$ and polarization of $83\%$. These differences, permitted us to study how the contribution of environmental noise impacted the spin damping results.

Under spin-damping both the signal and noise measured by the magnetometer are suppressed. This damping is characterized by the damping factor $DF$, or the loop gain of Fig.~\ref{Pp-MagSDSchematic}(b). The damping factor is determined with the feedback circuitry disconnected from the output of the magnetometer. A known voltage $V_1$ at the magnetometer resonance frequency is applied to the input of the feedback circuitry. Under the field produced by $V_1$, the magnetometer produces a signal which is recorded as $V_2$, and so the ratio of $V_2$ to $V_1$ is the open loop gain. The damping factor is adjusted using the variable attenuator/switch.

\section{\label{Theory}Theory - Spin-damping in an atomic magnetometer}

Signal is generated from the potassium atoms, of number $N$, whenever the net electron spin polarization
\begin{align}
\mathbf{P} \equiv \frac{\langle \mathbf{S} \rangle}{\hbar S} = \frac{1}{N \hbar S} \sum_{i=1}^{N} \langle \mathbf{S}_i \rangle
\end{align}
is misaligned from the magnetic field $\mathbf{B} = B_0 \hat{z}$.  The output of the balanced polarimeter, shown in Fig.~\ref{Pp-MagSDSchematic}, is directly proportional to the transverse polarization along the probe beam direction
\begin{align} \label{eq:Vout}
V_{out} & = \alpha P_x.
\end{align}
The proportionality constant $\alpha$ can be viewed as the product of $\alpha_G$, the gain due to the polarimetry circuitry, and $\alpha_\phi$ given by the rotation of the probe polarization by \cite{Happer-Jau-Walker-2010}
\begin{align}
\phi            & = \alpha_\phi P_x    \\
\alpha_\phi     & = \frac{1}{2} n l r_e c f \frac{(\nu - \nu_0)}{(\nu - \nu_0)^2 + (\Delta \nu/2)^2 }
\end{align}
for a probe of frequency $\nu$ close to $\nu_0$, the D$_1$ transition frequency.  In the expression for $\alpha_\phi$, $n$ is the K number density, $l$ the length of the cell along the probe direction, $r_e = \frac{e}{m_e c^2}$  the classical electron radius, $f = 1/3$ the D$_1$ oscillator strength, and $\Delta \nu$ is the optical full-width at half-max (FWHM) line width.

The electrical signal $V_{out}$ is recorded by the spectrometer.  During spin damping, a fraction of the signal is fed back to a set of electromagnetic coils, creating a radio-frequency magnetic field
\begin{align} \label{eq:Bfb}
B_{fb} = \beta V_{out},
\end{align}
where the direction is chosen so as to push $\langle \mathbf{S} \rangle$ back into alignment with the static field and the constant $\beta$ is controlled by the circuitry of the feedback circuit.

An analogy with the feedback of a finite-gain amplifier \cite{Horowitz-Hill-1989} can be made to our system if we look at the steady-state response of the system to a resonant radio-frequency field $\mathbf{B}_{{\rm RF}} = B_1 \cos \left( \omega_L t \right) \hat{y}$, where the K Larmor frequency is $\omega_L = \gamma_K B_0$ and $\gamma_K = 2 \pi \times 700$~kHz/G.
With this input, and assuming that the optical pumping rate along $\hat{z}$ is much larger than the nutation rate, the response of the K atoms along the probe direction is
\begin{align} \label{SSresponse}
P_x =  \frac{1}{2} P_z \gamma_K T_2 B_1 \cos \omega_L t.
\end{align}
With spin damping turned on, the fraction of this response returned to the input is $\alpha \beta $ as defined by Eqs.~\ref{eq:Vout} and \ref{eq:Bfb}. The transverse polarization is then reduced or damped to
\begin{align}
P_x = \frac{ P_z \gamma_{K} T_2/2}{1 + \alpha \beta P_z \gamma_K T_2/2} B_1 \cos \omega_L t.
\end{align}
In analogy with the finite gain amplifier, we therefore label $\alpha \beta  P_z \gamma_{K} T_2/2 $ as the open loop gain, or damping factor $DF$, and the quantity $\left(1 + DF \right)$ as the return difference; note the resonant signal amplitude is reduced by the return difference.

More generally, the response of the magnetometer under a nearly resonant field of $ B_1 \cos ( \omega t )\hat{y}$ turned on at time $t = 0$, is
\begin{align} \label{undamped}
P_x  & = \frac{1}{2} \left[ \frac{\frac{1}{2} \gamma_K T_2}{1 + i (\omega -\omega_L) T_2} \left(e^{i \omega t} - e^{ i \omega_L t} e^{-\frac{t}{T_2}}\right) P_z B_1 + (P_x^0 + i P_y^0) e^{ i \omega_L t} e^{-\frac{t}{T_2}} \right] + cc,
\end{align}
where $cc$ stands for the complex conjugate of the proceeding expression, and $P_x^0$ and $P_y^0$ represents the initial $x$ and $y$ polarization, respectively.
This expression, as is Eq.~\ref{SSresponse} for the resonant steady-state response, is derived in limit of high longitudinal polarization, using the optical Bloch equation \cite{Corney-1997} for the atomic angular momentum $\langle \mathbf{F} \rangle$, and taking $ \langle \mathbf{S} \rangle/S =  \langle \mathbf{F} \rangle/F$; it is equivalent to that found in Ref.~\cite{Appelt-Happer-etal-1998} under the same polarization limit.

In the presence of feedback, $\mathbf{B}_{fb} = - \hat{y} \alpha \beta P_x $ is applied to the magnetometer and an additional $T_2$ type relaxation term is added to the Bloch terms with a corresponding relaxation rate of  $ \alpha \beta P_z \gamma_K/2$.  Defining an effective relaxation rate $\frac{1}{T_{2d}} = \frac{1}{T_2}  + \frac{\alpha \beta P_z \gamma_K }{2} = \frac{1}{T_2}( 1 + DF)$, the response of the magnetometer is similar in appearance to Eq.~\ref{undamped},
\begin{align} \label{damped}
P_x & = \frac{1}{2} \left[ \frac{\frac{1}{2} \gamma_K T_{2d}}{1+i(\omega-\omega_L)T_{2d}} \left(  e^{i \omega t} - e^{i \omega_L t}e^{-\frac{t}{T_{2d}}}\right) P_z B_1 + (P_x^0 + i P_y^0) e^{i \omega_L   t} e^{-\frac{t}{T_{2d}}} \right] + cc  .
\end{align}
Therefore the effect of damping is to  increase the relaxation rate by the return difference $(1 + DF)$, resulting in a suppressed signal and quicker response time, or broadened bandwidth. For unwanted initial perturbations of the magnetometer, represented by $P_x^0$ and $P_y^0$, damping provides a way to quickly return the magnetometer to an aligned state, in preparation to detect the desired signal clearly.    We turn next to see how spin-damping effects noise in the system, and ultimately the sensitivity of the magnetometer.

Noise is added to the magnetometer at several places - environmental magnetic noise, light shift noise, and spin-projection noise add noise through the transverse polarization, photon shot noise adds noise through the balanced polarimeter, and instrumental noise is added through the amplification stage. The first three represent white noise which is colored through the detection by the magnetometer. The last two are white noise contributions under normal detection by the magnetometer but become colored under the presence of feedback.

\subsection{Magnetic noise}

We begin by determining the noise in the $x$-polarization under the effects of magnetic noise - either environmental noise or light shift noise masquerading as a fictitious magnetic field in the direction of the probe beam \cite{Savukov-Sauer-etal-2005}.  The noise power spectral density in $P_x$, or $\mathcal{S}_{Px}$, can be related to transverse polarization noise in a frame rotating with the Larmor frequency through
\begin{align} \label{PSDlabtorot}
\mathcal{S}_{Px}(\omega) = \frac{1}{4} \left[ \mathcal{S}_{Px^\prime}(\omega - \omega_L) + \mathcal{S}_{Py^\prime}(\omega - \omega_L) + \mathcal{S}_{Px^\prime}(\omega + \omega_L) + \mathcal{S}_{Py^\prime}(\omega + \omega_L) \right],
\end{align}
where the primed coordinates denote the rotating frame and are related to the unprimed coordinates through $ \hat{x} + i \hat{y} = (\hat{x}^\prime + i \hat{y}^\prime)e^{i \omega_L t} $.   Within the rotating frame, the Fourier transform of the Bloch equations give the relationship between the power spectral density of the transverse polarization to that of the magnetic noise,
\begin{align} \label{PSDPtoB}
\mathcal{S}_{Px^\prime}(\omega) + \mathcal{S}_{Py^\prime}(\omega) = \left| h(\omega) \right|^2 \left[  \mathcal{S}_{Bx^\prime}(\omega) + \mathcal{S}_{By^\prime}(\omega) \right],
\end{align}
where the transfer function is $h(\omega) = \frac{i \gamma_K}{i \omega + \frac{1}{T_2}}$.
Therefore, using Eqs.~\ref{PSDlabtorot} and \ref{PSDPtoB},
\begin{align}
\mathcal{S}_{Px}(\omega) &= \frac{1}{4} |h(\omega-\omega_L)|^2 \left[ \mathcal{S}_{B x^\prime}(\omega-\omega_L) + \mathcal{S}_{B y^\prime}(\omega-\omega_L) \right]  \notag \\
                         & + \frac{1}{4} |h(\omega+\omega_L)|^2 \left[ \mathcal{S}_{B x^\prime}(\omega+\omega_L) + \mathcal{S}_{B y^\prime}(\omega+\omega_L) \right].
\end{align}
In the limit that $\omega$ is close to $\omega_L$ and $\omega_L T_2 >> 1$, the second term on the right-hand side can be neglected.  In a similar manner to Eq.~\ref{PSDlabtorot}, we can relate the magnetic noise in the rotating frame to that of the lab frame and $\mathcal{S}_{P_x}$ can be simplified to
\begin{align} \label{mnoise}
\mathcal{S}_{Px}(\omega) &= \frac{1}{8} \frac{P_z^2 \gamma_K^2 T_2^2}{(\omega-\omega_L)^2 T_2^2 + 1} \left[  \mathcal{S}_{Bx}(\omega ) + \mathcal{S}_{By}(\omega) \right],
\end{align}
where we take as our convention a one-sided power spectral density \cite{Percival-Walden-1993}.

With the addition of spin-damping the transfer function changes to $h(\omega) = \frac{i \gamma_K}{i \omega + \frac{1}{T_{2d}}}$ and the noise power spectral density is the same as in Eq.~\ref{mnoise}, but with $T_2$ replaced with $T_{2d}$.  From Eq.~\ref{damped} and \ref{mnoise} the signal to noise ratio of the absorptive signal, under steady-state conditions and for long acquisition time $T$, is
\begin{align}
SNR(\omega) = \frac{ SNR^0 }{\sqrt{1+ (\omega-\omega_L)^2 T_{2d}^2}},
\end{align}
where $SNR^0 = \left( B_1 \sqrt{T} \right) \sqrt{\frac{1} {\mathcal{S}_B} }$ is the resonant SNR and $\mathcal{S}_B = \frac{1}{2} \left[ \mathcal{S}_{B_x}(\omega ) + \mathcal{S}_{B_y}(\omega) \right] $ represents the average magnetic noise in any given direction.  From this expression it is easy to see that the resonant SNR does not depend on damping and that the FWHM line width is
\begin{align}
\Delta \omega = \frac{\sqrt{12}}{T_{2d}} = \frac{(1+DF)\sqrt{12}}{T_{2}}.
\end{align}
Therefore the bandwidth of the sensitivity for an absorptive signal increases as the return difference, without loss of SNR, as long as the only noise is magnetic noise.

\subsection{Spin-projection noise}

We consider, at first, only a single potassium atom in the cell, but leave off the superscript $i$ for notational simplicity. As described by Ref.~\cite{Braun-Konig-2007}, the spin projection noise associated with measurement of $S_x$ can be calculated by
\begin{align}
\mathcal{S}_{Sx}(\omega) = 2 \times \int^0_{\infty}  R_{Sx}(t) \left( e^{-i \omega t} + e^{i \omega t} \right) dt,
\end{align}
where the symmetrized spin-spin autocorrelation function $R_{Sx}$ is given by
\begin{align} \label{ssautocorrelation}
R_{Sx}(t) =  \frac{1}{2} \mathrm{Tr} \left\{\rho(0) \left[S_{x}^H(t) S_{x}^H(0) + S_{x}^H(0)S_{x}^H(t) \right] \right\}.
\end{align}
In the above expression, $\rho(0)$ is the density matrix at time $t = 0$ and $S_{x}^H$ is the operator $S_x$ in the Heisenberg representation. In the absence of magnetic noise and in the limit of high polarization, the solution to the Bloch equation in the Larmor rotating frame and with damping gives
\begin{align} \label{evolution}
\langle S_x \rangle & = \mathrm{Tr} \left\{ \rho(0) \left[ S_x \cos \omega_L t - S_y \sin \omega_L t\right] e^{ - \frac{t}{T_2^\prime}} \right\} \\
                    & = \mathrm{Tr} \left\{ \rho(0) S_{x}^H(t) \right\}. \nonumber
\end{align}
Equation~\ref{evolution} implies that $S_{x}^H(t)$ can be replaced by $\left( S_x  \cos \omega_L t -  S_y  \sin \omega_L t\right) e^{ - \frac{t}{T_2^\prime}}$ in Eq.~\ref{ssautocorrelation}, in which case the spin-spin autocorrelation function becomes
\begin{align} \label{ssautocorrelation2}
R_{Sx}(t) = \mathrm{Tr} \left\{ \rho(0) \left[ S_{x}^2 \cos \omega_L \tau - \frac{1}{2}(S_{y} S_{x}  + S_{x}S_{y})\sin \omega_L \tau \right] e^{- \frac{\tau}{T_2^\prime}} \right\} .
\end{align}
Therefore the average power spectral density per atom is
\begin{align} \label{spinprojectionnoise}
\mathcal{S}_{Sx}(\omega)  = \frac{\hbar^2}{2} \frac{T_{2d}}{1 + T_{2d}^2 (\omega - \omega_L)^2 },
\end{align}
where we have taken the limit that $\omega$ is close to $\omega_L$ and $\omega_L T_2 >> 1$.  This agrees with quantum mechanical expression of Ref.~\cite{Braun-Konig-2007} derived for a spin-1/2 particle.  While it is clear from Eq.~\ref{spinprojectionnoise} that the resonant noise density is reduced with spin-damping the net power is not.  Therefore this reduction would not be considered spin-squeezing, rather it represents the broadening of the spectrum; nevertheless the ability to easily vary resonant noise and width may be of use in quantum control.

The noise power spectral density for the net magnetic moment along the probe direction is related to $\mathcal{S}_{Fx}(\omega)$ of Eq.~\ref{spinprojectionnoise} by
\begin{align} \label{spinprojectionnoise2}
\mathcal{S}_{Px} = 4 \frac{\mathcal{S}_{Sx}(\omega)}{ \hbar^2 N}.
\end{align}
Therefore the SNR under spin-projection noise is determined by Eqs.~\ref{damped} and \ref{spinprojectionnoise2}:
\begin{align} \label{SNRspinprojection}
SNR = \frac{SNR^0}{\sqrt{ 1 + (\omega -\omega_L)^2 T_{2d}^2}} \sqrt{\frac{ T_{2d}}{ T_2 } },
\end{align}
where the resonant undamped $SNR^0$ is
\begin{align} \label{SNR0}
SNR^0 = B_1 \sqrt{T} \left[ P_z \gamma_K \sqrt{\frac{N  T_{2}}{8} } \right].
\end{align}
Note the inverse of the square bracketed expression in Eq.~\ref{SNR0} represents the undamped resonant field sensitivity, or the spin-projection noise expressed in terms of field.  From Eq.~\ref{SNRspinprojection}, the SNR bandwidth increases as $\Delta \omega = \frac{(1+DF)\sqrt{12}}{T_{2}}$, as in the case of magnetic noise.  However unlike the case of magnetic noise, this broadening comes at a cost to SNR; the resonant SNR decreases in proportion to $\sqrt{1+DF}$.

\subsection{Photon shot noise}

Through interaction with the K atoms, the polarization angle of the probe beam after the magnetometer $\phi$ is shifted from its original phase $\phi_0$ by $\phi = \phi_0 + \alpha_\phi P_x$.  During feedback, using the optical Bloch equations, and in the limit that $\omega$ is close to $\omega_L$ and $\omega_L T_2 >> 1$, the Fourier transform of $\phi$ is equal to the transform of $\phi_0$ times the transfer function $h(\omega) = \frac{\frac{1}{T_2} + i (\omega - \omega_L)}{\frac{1}{T_{2d}} + i (\omega - \omega_L)}$.  Therefore the power spectral density of $\phi$ is
\begin{align} \label{photonshotnoise}
\mathcal{S}_\phi(\omega) = \left[ \frac{ 1 + (\omega - \omega_L)^2 T_{2}^2}{1 + (\omega - \omega_L)^2 T_{2d}^2}  \right] \frac{T_{2d}^2}{T_2^2} \mathcal{S}_{psn}(\omega),
\end{align}
where $\mathcal{S}_{psn}$ is the standard white photon shot noise.

Therefore the SNR from photon shot noise alone can be expressed as
\begin{align} \label{SNRpsn}
SNR  = SNR^0 \sqrt{\frac{1}{[1 + (\omega- \omega_L)^2 T_2^2  ][1 + (\omega- \omega_L)^2 T_{2d}^2  ]}},
\end{align}
where the resonant SNR under no damping is given by
\begin{align}
SNR^0 = B_1 \sqrt{T} \left[ \frac{ P_z \gamma_K T_2 \alpha_\phi/2} {\sqrt{ \mathcal{S}_{psn}}} \right] .
\end{align}
From Eq.~\ref{SNRpsn} it easy to see that the resonant SNR does not change with damping, but  the FWHM linewidth of this SNR modestly increases from $\frac{2}{ T_2}$ with no damping to $\frac{\sqrt{12}}{T_2}$ for infinite damping, with most of the increase occurring for damping factors under 10.

\subsection{Total noise and bandwidth}

The measurement of the noise under spin-damping in principle permits the identification of the separate contributions of spin-projection noise $\mathcal{S}_S = \alpha^2 \mathcal{S}_{P_x}$ from Eq.~\ref{spinprojectionnoise2}, photon shot noise $\mathcal{S}_P = \alpha_G^2 \mathcal{S}_{\phi}$ of Eq.~\ref{photonshotnoise}, and magnetic noise $\mathcal{S}_{B} = \alpha^2 \mathcal{S}_{P_x}$ of Eq.~\ref{mnoise}.  The total magnetometer noise power spectral density can be expressed as
\begin{align} \label{TotalNoisePD}
\mathcal{S}_V(\omega)   & \equiv \mathcal{S}_S(\omega) + \mathcal{S}_P(\omega) + \mathcal{S}_B(\omega) \nonumber \\
                        & = \frac{A_n^2 + (\omega - \omega_L)^2 T_{2d}^2 B_n^2 } {1 + (\omega - \omega_L)^2 T_{2d}^2},
\end{align}
where in the second expression the functional dependence on $\omega$ has been made explicit such that $A_n^2$ represents the amplitude on resonance and $B_n^2$ the base noise at large off-resonance values.  The two parameters $A_n^2$ and $B_n^2$ can be expressed in terms of the resonant noise spectral densities with no damping applied, denoted in the following by a zero superscript,
\begin{align}
B_n^2 & = \mathcal{S}_P^0 \\
A_n^2 & = a x^2 + b x = (\mathcal{S}_P^0 + \mathcal{S}_B^0) x^2 + \mathcal{S}_S^0 x, \label{ResonantAmplitude}
\end{align}
where $x \equiv \frac{T_{2d}}{T_2} = \frac{1}{(1+DF)}$.  If in addition, to these noise sources, there is an out-of-loop noise source, say for instance from the spectrometer itself, both base noise power $B_n^2$ and the amplitude noise $A_n^2$ would be increased by this constant noise.

The SNR under the combined noise can be found using Eq.~\ref{TotalNoisePD}.  Both the loss of SNR and the broadening of the SNR with spin damping depend on the relative amounts of the different types of noise.  In our experimental case where magnetic noise and photon shot noise dominate over spin-projection noise, we find that broadening with little loss of SNR can occur for damping factors on the order of 10 or less.

\subsection{Measuring noise}

For a finite acquisition time $T$ of the noise signal $V(t)$, the ensemble average of the periodogram $\left[\mathfrak{P}_T(\omega) \right]$ can be taken as a measurement of the frequency distribution of the noise \cite{Brown-Hwang-1997}
\begin{align} \label{AVperiodogram}
\left[\mathfrak{P}_T(\omega) \right] & \equiv \frac{1}{T} \left[|\mathcal{F}\{V(t)\}|^2 \right] \\
                                     & = \int^T_{-T}  \left(1 - \frac{|\tau|}{T} \right) R(\tau) e^{-i \omega \tau} d\tau,
\end{align}
where $\mathcal{F}\{V(t)\}$ is the Fourier transform and $R(\tau)$ is the autocorrelation function of $V(t)$.  In the limit that $T$ is much larger than the characteristic decay time of $R(\tau)$ with $\tau$, $\left[\mathfrak{P}_T(\omega) \right]$ approaches half the power spectral density, $\frac{1}{2} \mathcal{S}(\omega)$.    More generally, the integral on the right hand side of Eq.~\ref{AVperiodogram} can be viewed as the Fourier transform of the autocorrelation function multiplied by a triangular function, or the power spectral density convoluted with the function $T \mathrm{sinc}^2(\omega T/(2 \pi))$. For finite acquisitions times, the features of the power spectral density are broadened on the order of $\frac{1}{T}$ to give $\left[\mathfrak{P}_T(\omega) \right]$.

In this paper we focus on the absorptive signal part of the signal as measured through quadrature detection, as is typically used in magnetic resonance techniques.  Such phase-sensitive detection is needed for an optimal signal to noise ratio and to distinguish the true signal from interfering signals.  The noise spectra for absorptive signals is $\sqrt{\frac{1}{2}\left[\mathfrak{P}_T(\omega) \right]  }$ and therefore for long $T$ is equivalent to
$\frac{1}{2}\sqrt{\mathcal{S}(\omega)}$. In the next section, however, the presented noise data is scaled so as to represent $\sqrt{\mathcal{S}(\omega)}$ for ease of comparison with derived expressions for noise spectral density.

\section{\label{Results}Results}

\subsection{Spin-damping at long times}

\begin{figure}[tb]
\centerline{\includegraphics[width=6in]{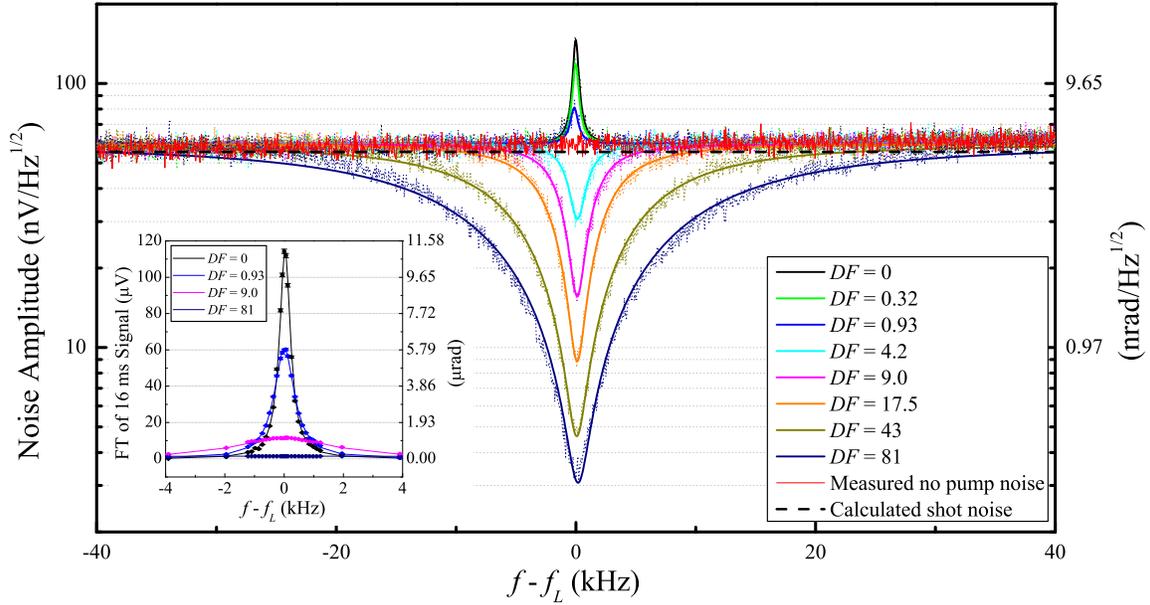}}
\caption[Noise Suppression]{Spectra of measured magnetometer noise (dotted lines) for various damping factors, showing suppression of on-resonance noise amplitude $A_n$ to well below photon shot noise at the higher damping factors. The noise spectra for each damping factor is fitted (solid lines) to Eq.~\ref{TotalNoisePD}. Also plotted are the measured noise (red) with no pump beam and the expected photon shot noise (black dashed line). The inset shows similar suppression of the magnetometer output to a reference RF magnetic signal of $209~\rm fT$ (points), fitted to a Lorentzian function. Acquisition time was 16.4~ms, more than an order of magnitude larger than K $T_2$, and this data is expressed both in volts output from the polarimeter, left axis, and rotation angle of the probe polarization, right axis. }
\label{Po-NoiseSigMay24}
\end{figure}

\begin{figure}[tb]
\centerline{\includegraphics[width=6in]{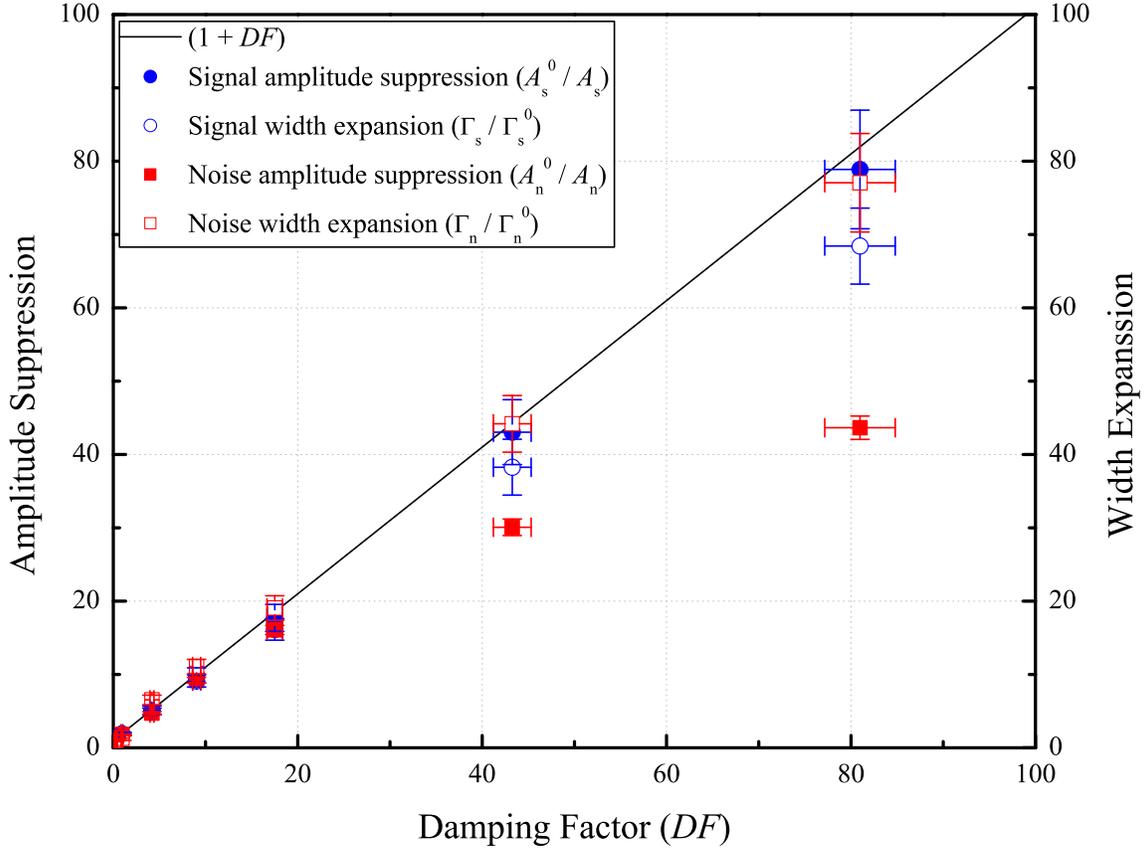}}
\caption[Noise suppression]{The magnetometer signal amplitude suppression, defined as $A_s^0/A_s$ (solid blue), and linewidth expansion, defined as $\Gamma_s/\Gamma_s^0$ (open blue), scale as $(1~+~DF)$ (black solid line). The on-resonance noise amplitude suppression, $A_n^0/A_n$ (solid red), is linear at low damping factors ($<20$) but clearly reaches a noise limit at higher $DF$, while the expansion of noise width, $\Gamma_n/\Gamma_n^0$ (open red), is linear for the range of damping factors measured.}
\label{Po-SuppressionMay24}
\end{figure}

When spin-damping is applied to the magnetometer, both the signal and noise are suppressed when resonant with the Larmor frequency of the magnetometer and their effective widths are broadened, as shown in Fig.~\ref{Po-NoiseSigMay24} for cell 1.  For the absorptive signal the resonant amplitude $A_s$ is inversely proportional to the return difference $(1+ DF)$, while the FWHM width $\Gamma_s = \frac{1}{\pi T_{2d}}$ is proportional to the return difference. This is clearly demonstrated in Fig.~\ref{Po-SuppressionMay24}, where $A_s$ and $\Gamma_s$ are determined from fits of the signal-versus-frequency data to a Lorentzian function, the form expected from Eq.~\ref{damped}. For clarity, the parameters in Fig.~\ref{Po-SuppressionMay24} have been normalized with respect to their undamped counterparts $A_s^0$ and $\Gamma_s^0$.  Representative absorptive signals and fits for select damping factors are shown in the inset of Fig.~\ref{Po-NoiseSigMay24}.

As evident in Fig.~\ref{Po-NoiseSigMay24}, the noise spectra have a very different functional form from the signal. Fits of these spectra to the square root of Eq.~\ref{TotalNoisePD} are depicted as solid lines.  Good agreement between the data and fits are observed, except for the highest damping factors, where the fit slightly underestimates the on-resonance amplitude and deviates from the high off-resonance frequency behavior. The width of the noise peak/dip is predicted by Eq.~\ref{TotalNoisePD} to be equal to $\Gamma_n$, which is equivalent to $\Gamma_s$, and therefore should increase as the return difference.  The fit parameter $\Gamma_n$ demonstrates this predicted behavior in Fig.~\ref{Po-SuppressionMay24}. In contrast, the resonant noise amplitude $A_n$ is suppressed as the signal for low damping factors, but is suppressed less than the signal at higher damping factors.  As explored more below, this behavior is expected from Eq.~\ref{ResonantAmplitude}. The third parameter, $B_n$, predicts the far off-resonant amplitude of the noise corresponding to the photon shot noise. The slight increase in the measured shot noise ($\simeq~25~\rm nV/Hz^{1/2}$) is due to additional observed noise from the balanced polarimeter as is measured in the absence of both probe and pump light.

\begin{figure}[tb]
\centerline{\includegraphics[width=6in]{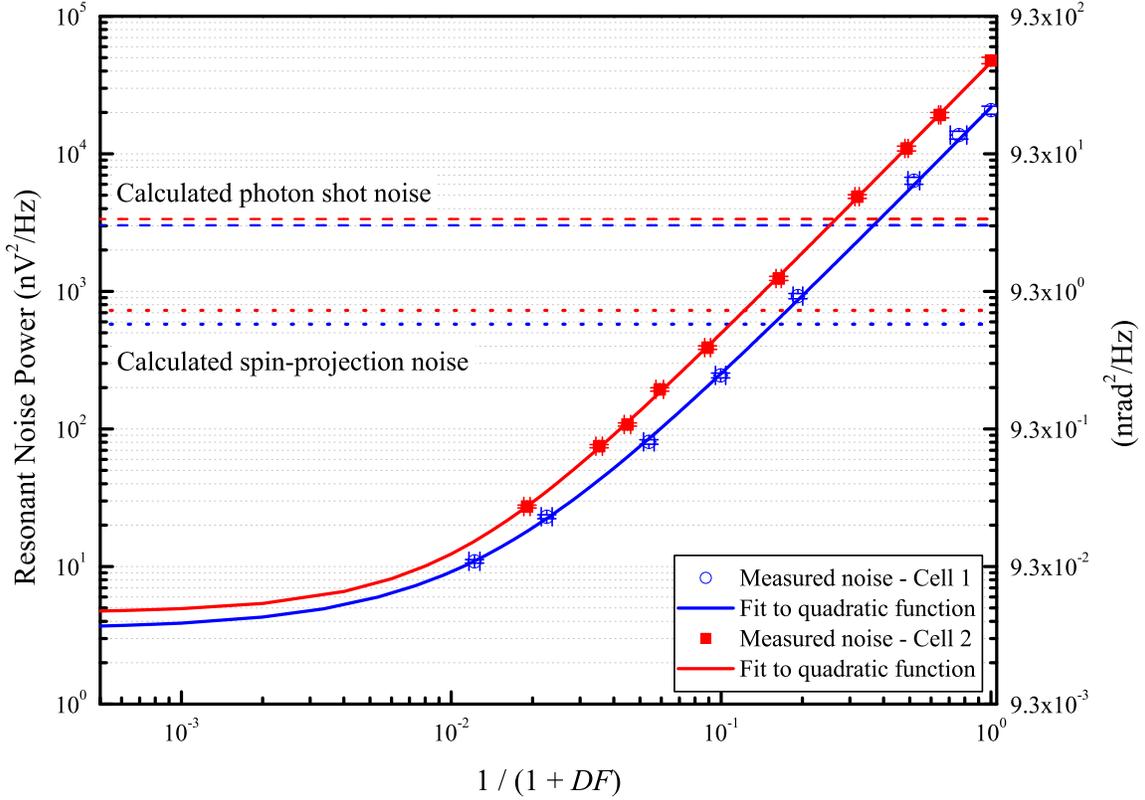}}
\caption[Quadratic fit of the suppressed noise amplitudes]{For cell 1 (open blue) and cell 2 (solid red), the plot of resonant noise power $A_n^2$ as a function of suppression, fitted to a quadratic polynomial (lines). The values of all parameters for the two cells are given in Table~\ref{NoiseTable}.}
\label{Po-NoisePower1}
\end{figure}

For comparison, a set of noise measurements are made with a second K cell, cell 2, which operated with a higher level of environmental noise. For both cells, the resonant noise power $A_n^2$  is plotted in Fig.~\ref{Po-NoisePower1} as a function of $x = \frac{1}{1+DF}$. The measured noise power is fitted to a quadratic polynomial of the form $ax^2+bx+c$, corresponding to Eq.~\ref{ResonantAmplitude}.  From the fit we extract the noise contributions, with $a$ corresponding to  $\mathcal{S}_P^0 + \mathcal{S}_B^0$, and $b$, to the spin-projection noise. Parameter $c$ represents the limit of noise suppression, and may be due to external noise added outside of the feedback loop or noise folded back into the spectrum from aliasing effects and limitations in the spectrometer's filtering. This noise power is more than an order of magnitude larger than the noise floor of the spectrometer.

The values of the fit parameters for both data sets are given in Table~\ref{NoiseTable}. The measured spin-projection noise is similar in magnitude and agrees fairly well with the predicted values. While the calculated noise takes into consideration the reduced polarizations of 83\% and 78\% for cells 1 and 2 respectively, the derivation of Eq.~\ref{SNR0} relies heavily on the atomic system being in the high polarization limit; this may be responsible for the observed trend that the predicted noise is higher than measured noise, particularly for the lower polarization cell. The quoted errors for the calculated spin-projection noise are due to the uncertainty in the parameters $n_{\rm K}$, $P_z$, $T_2$, and the volume of the cell. Fig.~\ref{Po-NoisePower1} clearly shows that at high damping factors, we are able to suppress the total magnetometer noise power by about three order of magnitude below photon shot noise and two orders of magnitude below the undamped spin-projection noise.

\begin{table}
\caption{Using the fit parameters from the data in Fig. 4, the wings of the noise curve $B_n$, and the magnetometer responsivity, we find the resonant noise contributions. The measured and predicted values for spin-projection noise are in reasonable agreement. For the magnetic noise, which is predominantly environmental, we give as a predicted lower bound the calculated light shift noise.  The measured shot noise is close to the predicted value.  For the prediction of the out-of-loop noise we only give a lower bound corresponding to the base noise of the spectrometer itself; aliasing effects may account for the observed noise.}
\centering
  \begin{tabular}{| c | c | c | c | c | c |}
    \hline
    \multicolumn{2}{|c|}{\multirow{3}{*}{\textbf{Fit Parameters}}}      & \textbf{Magnetic Noise}               & \textbf{Photon Shot}              & \textbf{Spin-projection}      & \textbf{Out-of-loop}          \\
    \multicolumn{2}{|c|}{}                                              & \textbf{$\sqrt{a - B_n^2}$}           & \textbf{noise, $B_n$}             & \textbf{noise, $\sqrt{b}$}    & \textbf{noise, $\sqrt{c}$}    \\
    \multicolumn{2}{|c|}{}                                              & ($\rm{aT/}\sqrt{\rm Hz}$)             & ($\rm{aT/}\sqrt{\rm Hz}$)         & ($\rm{aT/}\sqrt{\rm Hz}$)     & ($\rm{aT/}\sqrt{\rm Hz}$)     \\
    \hline
    \multirow{2}{*}{\textbf{Cell 1}}                    & Measured      &  $248\pm19$                           & $107\pm7$                         & $35\pm5$                      & $3\pm1$                       \\
    \cline{2-5}
                                                        & Predicted     &  $>2\pm1$                             & $100\pm2$                         & $45\pm9$                      & $>0.4$                        \\
    \hline
    \multirow{2}{*}{\textbf{Cell 2}}                    & Measured      &  $361\pm28$                           & $118\pm8$                         & $31\pm5$                      & $4\pm1$                       \\
    \cline{2-5}
                                                        & Predicted     &  $>2\pm1$                             & $105\pm2$                         & $47\pm9$                      & $>1.0$                        \\
    \hline
  \end{tabular}
  \label{NoiseTable}
\end{table}

\begin{figure}[h!]
\centerline{\includegraphics[width=6in]{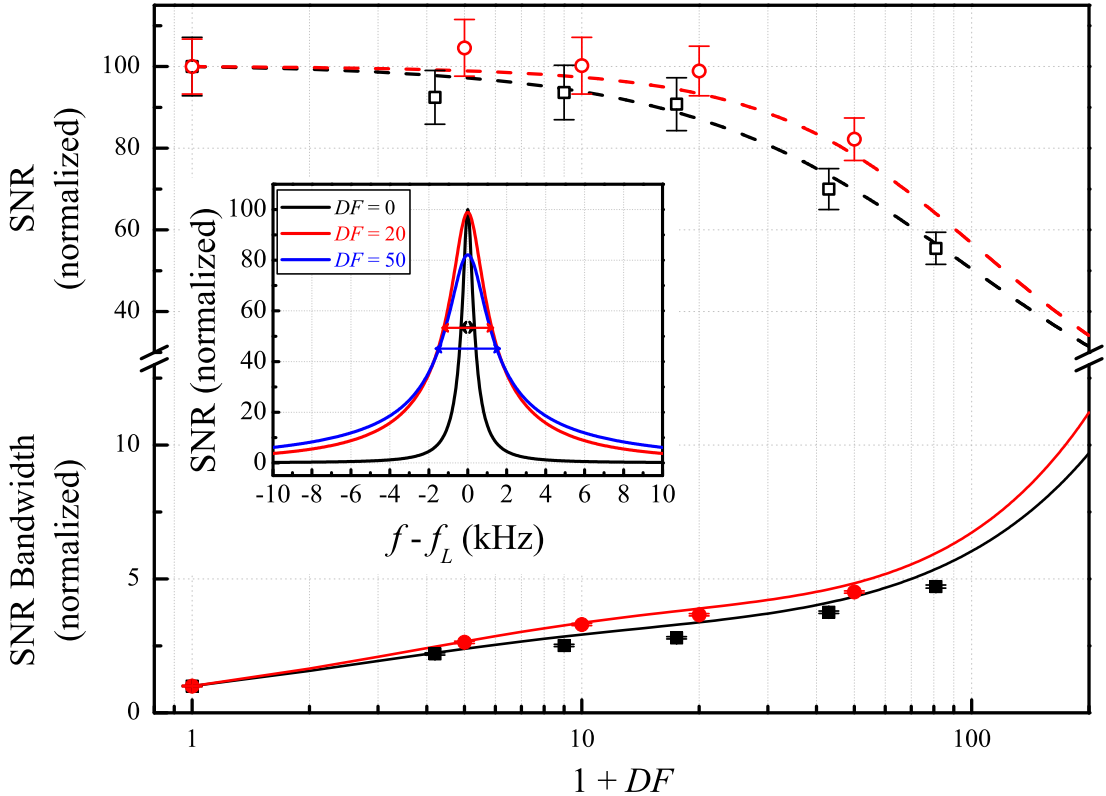}}
\caption[Magnetometer SNR bandwidth with spin-damping]{Plot of the measured magnetometer $SNR$ (open points) and bandwidth (solid points), for both cell 1 (black square) and cell 2 (red circles), shows good agreement to the predicted (lines) values. The inset is the magnetometer $SNR$ for three damping factors as a function of frequency for  cell 2, and shows that sensitivity bandwidth can be broadened with little loss of SNR for $DF \leq 20$.}
\label{SDo-SNR}
\end{figure}

The SNR of the magnetometer is simply calculated by taking the ratio of the fit equations corresponding to the measured signal and the noise for each $DF$. Figure~\ref{SDo-SNR}, shows that as the damping is increased the SNR bandwidth, or sensitivity bandwidth, increases, but at the same time the resonant SNR decreases.  For damping factors of about $20$ or less, however, this loss of signal is quite small, so that broadening of the bandwidth in this regime comes with little cost.  For cell 1, a damping factor of 17.5 increases the detection bandwidth of the magnetometer by a factor of 2.8 over 0.70~kHz with $\sim10\%$ loss in on-resonance sensitivity, while cell 2 shows a bandwidth increase of $3.7\times$ over 0.74~kHz with almost no loss in sensitivity for $DF = 20$. The difference between the two data sets can be mostly attributed to the higher level of environmental noise experienced by cell~2 compared to cell~1.

The increase in bandwidth in an atomic magnetometer can significantly reduce the detection time when the frequency of the signal to be detected is not well known. For example, in NQR detection the resonant frequency of the material is temperature dependent; our test substance has a temperature coefficient of $100~\rm Hz/^\circ C$. Therefore, a factor of 3 increase in bandwidth without loss in sensitivity is equivalent to a factor of 3 increase in the acceptable temperature variation of the substance under detection.

\subsection{Spin-damping at short times and in the presence of ringing}

\begin{figure}[h!]
\centerline{\includegraphics[width=6in]{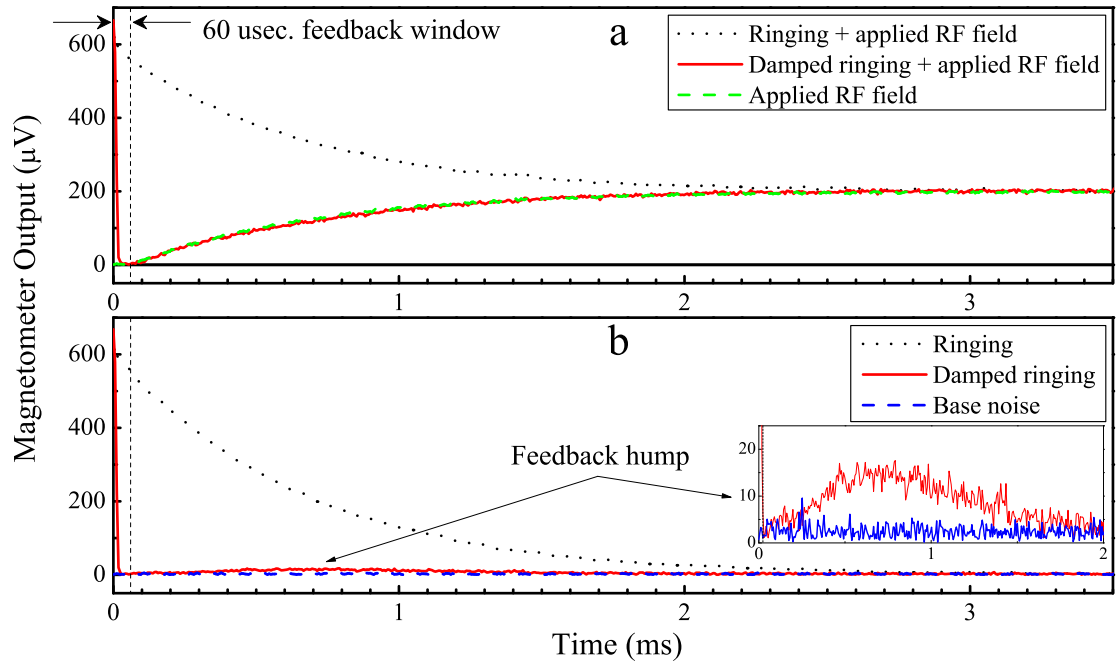}}
\caption[Ringing and signal with and without spin-damping]{Plots (a) and (b) demonstrate the application of spin-damping with the magnetometer initially in a perturbed state, a state created by a RF pulse of amplitude $1.13~\rm pT$ applied for $t\leq0$. (a) The magnetometer response to a $0.37~\rm pT$ RF signal, applied at $t=60~\rm \mu s$, is obscured by the transient ringing from the initial perturbed state (dotted line).  Application of spin-damping during a short window quickly eliminates the transient and permits the clear observation of the signal (solid line) as compared to when there is no initial perturbation (dashed line). (b) The ringing transient naturally decays with a time constant of $T_2= 0.7~\rm ms$  (dotted line), but under damping decays in less than $60~\rm \mu s$ (solid line).  However, a small feedback hump, arises after the damping field is turned off, due to inhomogeneity in $B_0$ across the K cell.}
\label{SDo-RingingSignalTimeOct18}
\end{figure}

\begin{figure}[h!]
\centerline{\includegraphics[width=6in]{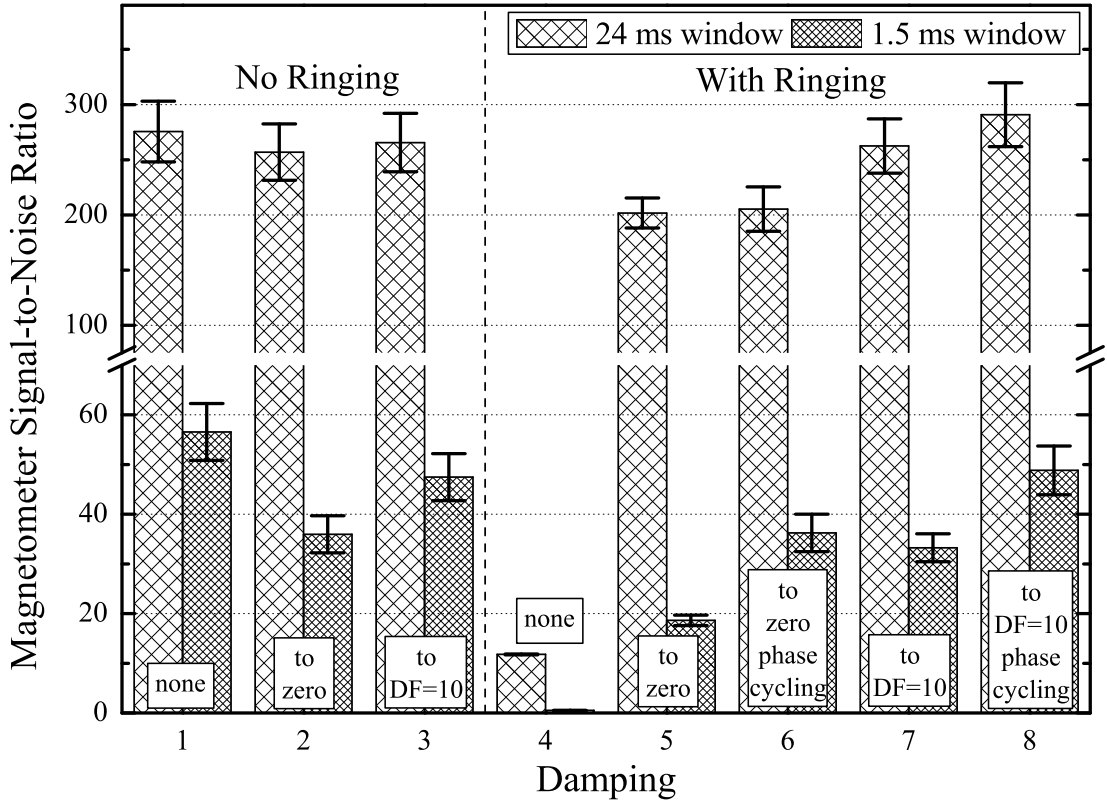}}
\caption[Magnetometer SNR in the presence of ringing and spin-damping]{Results showing that application of spin-damping can reduce the negative effects of perturbation noise and recover the magnetometer sensitivity. Two forms of damping are tested. Both start with $DF = 150$ for the first $20~\rm \mu s$ of the $60~\rm \mu s$ feedback window, but in one the DF is smoothly reduced to zero, while in the other the DF is reduce to $10$ and kept at this value throughout the acquisition window. SNR is measured for a $24~\rm ms$ (sparse hatching) and a $1.5~\rm ms$ (dense hatching) window in the absence, columns (1)-(3), and in the presence, (4)-(8) of ringing created by a perturbing pulse three times larger than the detected signal. As shown in (1)-(3), the switching off of damping adds noise, but with damping retained during acquisition the SNR is regained. Measurement 4 shows the significant loss of SNR due to noise from an initial perturbation of the K spins.  The SNR is partially regained with damping (5).  The addition of phase cycling (6) or damping during the window (7) further increases the SNR, and with the combination of the two techniques (8) the SNR for both window sizes is in agreement with the SNR when ringing is not present (1).}
\label{Po-SNRNov11}
\end{figure}

Any net magnetization transverse to $B_0$ has a ring down with the time constant $T_2$.  If such a component exists at the beginning of a measurement the associated ringing can dwarf the signal of interest, as demonstrated in Fig.~\ref{SDo-RingingSignalTimeOct18}. This is particularly detrimental for short data acquisition times or short-lived signals, as shown in Fig.~\ref{Po-SNRNov11}. To illustrate the potentially catastrophic effects of ringing in a high $Q$ atomic magnetometer we apply a long perturbing pulse ending at time $t=0$, the beginning of the data acquisition windows of Fig.~\ref{SDo-RingingSignalTimeOct18}. During the first millisecond in Fig.~\ref{SDo-RingingSignalTimeOct18}(a), the ringing clearly masks the desired signal, in this case a three times smaller radio-frequency pulse applied at $t = 60~\rm \mu s$.

The application of spin damping in the first $60~\rm \mu s$ permits for the quick damping of the ringing and clear detection of the desired signal, shown as a solid line in Fig.~\ref{SDo-RingingSignalTimeOct18}(a).
Figure~\ref{SDo-RingingSignalTimeOct18}(b) shows that the ringing decay constant is reduced by approximately a factor of 50 under the effects of damping.  In both figures, the negative feedback starts at a high damping factor of $\sim150$ for approximately the first $20~\rm \mu s$ and is smoothly ramped down to $DF=0$ over the following $40~\rm \mu s$, so as to avoid the creation of undesirable transients from the turn-off of damping.

There is, however, a small rise in the magnetometer signal following the application of feedback; the arrow in Fig.~\ref{SDo-RingingSignalTimeOct18}(b) indicates the emergence of this ``feedback hump." Through modeling, it is determined that this small rise is due to the inhomogeneity in $B_0$ across the K cell. The applied feedback field forces the net magnetic moment of the cell to zero. Some isochromats across the cell become $180^\circ$ out of phase from one another and once damping is off, individual isochromats with different Larmor frequencies partially rephase and a small magnetic moment re-emerges. For measurements in which the phase of the signal can be controlled separately from the perturbation, as is common for echo experiments in magnetic resonance, flipping or cycling the phase of the signal can be used to cancel the effects of this feedback hump. Such phase cycling is  commonly used to suppress the effect of the transients created by the refocusing pulse.  The ameliorating impact of phase cycling is shown in Fig.~\ref{Po-SNRNov11}, through comparing the SNR data of columns (5) and (7) to columns (6) and (8), respectively.  The combination of spin-damping and phase-cycling together leads to a strong and rapid  suppression of the transients, at the same time helping to avoid saturation and a potentially long recovery time of the spectrometer.  Furthermore, the use of an atomic magnetometer for detection, permits the use of a low-$Q$ probe for excitation thus preventing long-time ringing of the excitation coil.

In addition to the coherent transient added by the feedback hump, the turning on and off of spin-damping adds noise to the magnetic field detection, even when the magnetometer begins in an aligned state. This noise can be greatly reduced, but not eliminated, by shaping the spin-damping to turn-off gently as was done for the data in Fig.~\ref{SDo-RingingSignalTimeOct18}. By comparing the SNR of a signal acquired without damping, column (1) of Fig.~\ref{Po-SNRNov11}, to SNR with damping applied before data acquisition, column (2), we can see that the loss of SNR is particularly evident for data acquisition over short times.  Note the shorter window associated with column (1) has a SNR that is nearly a factor of 5 smaller than the larger window, a result consistent with theoretical predictions.

One way to avoid the noise associated with switching damping off is to leave damping on during data acquisition. As discussed in the previous section, this can be done for low damping factors without loss of signal and with an increase in sensitivity bandwidth. The benefits to SNR can be clearly observed in Fig.~\ref{Po-SNRNov11}, by comparing columns (2) and (5) where there is no damping in the window, to columns (3) and (7) where damping, $DF = 10$, is left on during the window. Combining both phase-cycling and damping during acquisition, permits us to retain the sensitivity of the magnetometer even in the presence of ringing, Fig.~\ref{Po-SNRNov11} column (1) to column (8).  Therefore, and particularly for short windows as is necessary in magnetic resonance echo trains, it is important to have both continuance of damping into the window to avoid switching noise and the use of phase cycling to minimize the feedback hump. Armed with both these tools, spin-damping promises to be quite useful in the reduction of unwanted delay, or dead-time, before data acquisition.

\section{\label{Conclusion}Conclusion}

In this work, we have demonstrated that negative magnetic feedback can effectively be used to rapidly damp the ringing of the K spins from some unwanted initial perturbation.  Under spin-damping the effective $T_2$ can be reduced by more than an order of magnitude, therefore permitting the clear observation of short-lived signals, which otherwise would be obscured by the use of a high $Q$ atomic magnetometer.

Furthermore we find that the magnetometer suppresses not only coherent signals, but also noise. Damping effects the spectrum of the noise, both amplitude and shape, according to the type of noise, so that we are able to separately measure magnetic, photon shot, and spin-projection noise.  While the net power in the magnetic and photon shot noise are reduced under damping, the power in spin-projection noise remains the same even as its spectrum is broadened.  The magnetic noise spectrum also broadens, with the effective $T_2$ simply replacing the undamped $T_2$ in the spectral shape.  The photon shot noise, however, becomes colored under the presence of negative feedback, giving the noise spectrum an inverted appearance.  In total we observe a resonant noise an order of magnitude lower than the undamped photon shot noise, implying the closed-loop production of polarization-squeezed light.

For phase-sensitive detection, the signal and noise are broadened under damping so as to increase the bandwidth of the magnetometer.  For magnetic and photon shot noise, this increase is not accompanied by loss of SNR, while for spin-projection noise the resonant SNR decreases as the square root of the effective $T_2$. Therefore in our system, which is dominated by magnetic and photon shot noise, we observed a three times increase in detection bandwidth with little degradation to the sub-femtoTesla sensitivity of the magnetometer.

\begin{acknowledgments}
We would like to acknowledge Philip Naudus for his modeling of the spin system under damping and field inhomogeneity. This work was supported by NSF grants \#0730473 and \#054798.
\end{acknowledgments}

\bibliography{SpinDampingBib}

\end{document}